\documentclass[prl,floatfix,twocolumn,showpacs,amsmath,amssymb]{revtex4}

\usepackage{graphicx}
\usepackage{dcolumn}
\usepackage{bm}

\begin{document}
\title{Calculation of ${\boldsymbol\Delta} {\bf (k,}
{\boldsymbol\omega}{\bf )}$ for a 2D t-J Cluster}

\author{Didier~Poilblanc}
\email{Didier.Poilblanc@irsamc.ups-tlse.fr}
\affiliation{Groupe de Physique Th\'eorique\\
Laboratoire de Physique Quantique, UMR--CNRS 5626\\
Universit\'e Paul Sabatier, F-31062 Toulouse, France} 
\homepage{http://ww3-phystheo.ups-tlse.fr/~didier}

\author{D. J. Scalapino}
\affiliation{Department of Physics, University of California\\ 
Santa Barbara, California 93106}

\date{\today}

\begin{abstract}

Using numerical techniques, the diagonal and off-diagonal superconducting
one-electron Green's functions are calculated for a two-dimensional (2D)
t-J model on a periodic 32-site cluster at low doping.
From these Green's functions, the momentum and frequency dependence of the
pairing gap $\Delta ({\bf k}, \omega)$ are extracted.  It has $d_{x^2-y^2}$
symmetry and exhibits $\omega$-dependent structure which depend upon J/t.
We find that the pairing gap persists down to small J/t values. The
frequency- and momentum-dependent renormalized energy and renormalization
factor are also calculated.

\end{abstract}

\maketitle


For the last decade, the search for 
superconductivity in models of
strongly-correlated fermions, has been
triggered by accumulating experimental
evidence in favor of an unconventional
(i.e.~not based on the usual phonon-mediated
interaction) mechanism in high-$T_c$
two-dimensional (2D) \cite{TK00} or ladder
\cite{DR96} superconducting cuprates.  Soon after the
discovery of the layered cuprates, it was
proposed that the Hubbard model \cite{And87}
and its
strong coupling limit, the t-J model
\cite{ZR88}, 
captured the generic features of these
materials.  Early RPA \cite{MSV86,SLH86}
and Gutzwiller variational calculations
\cite{GJR87} supported the notion that
the doped state would have $d_{x^2-y^2}$
pairing. Nevertheless, despite the
conceptual simplicity of these
Hamiltonians and the results of these
approximate calculations, the nature
of the basic mechanism responsible for
pairing as well as the actual physical
properties of these models remained
controvertial.  In recent years, numerical
calculations have provided insight into the
second of these questions regarding the
actual properties of these models.
Specifically, Monte Carlo calculations
showed that the ground states of the
undoped models had long-range
antiferromagnetic order \cite{HT89,RRY89}. 
Exact
diagonalization studies showed that
holes doped into a
t-J cluster can form $d_{x^2-y^2}$ pairs
\cite{Poi93,Leu00}.  In addition,
numerical calculations support the view
that the doped system can have low-lying
stripe domain wall states \cite{SW00}
which can be stabilized by lattice
anisotropies \cite{KSW01,BCS01}.  While
the static striped phase competes with
superconductivity, it has been argued that
the addition of a next-near-neighbor
hopping \cite{WS99}
 or the use of periodic boundary
conditions \cite{Sor01} tips the balance
in favor of the $d_{x^2-y^2}$ pairing
phase.  The point is that a variety of
numerical calculations provide evidence
that indeed these models do exhibit the
basic properties seen in the cuprates.
However, in spite of
this progress, the question regarding the
nature of the basic mechanism responsible
for the pairing remains open.  Within the
traditional BCS framework, one would look
for a reflection of the pairing interaction
by examining the momentum and frequency
dependence of the gap $\Delta ({\bf k}, \omega)$.
Because of the relatively short coherence
length and the relatively high energy scales
J and t, one can hope to learn about
both the momentum and frequency
dependence of 
$\Delta ({\bf k}, \omega)$ from an exact
diagonalization study of a cluster.  Here we
present a numerical study aimed at
doing this for the t-J model and obtain
the first results for the
$\omega$-dependence of the gap.

\begin{figure}
\vspace{2mm} 
\includegraphics[width=0.45\textwidth]{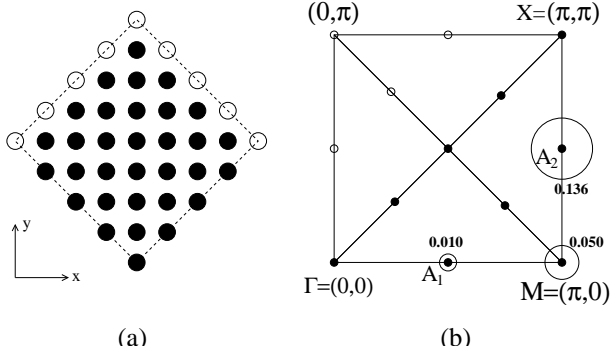}
\caption{\label{fig:cluster}
(a) Periodic square 32-site cluster; (b) Reciprocal space of the 32-site
cluster. The area of the circles (drawn only for a few 
${\bf k}$-points especially relevant to our analysis) correspond to
the {\it hole} occupancies 
$\big<{\tilde c}_{{\bf k},\sigma}{\tilde c}_{{\bf k},\sigma}^\dagger\big>$
in the 2-hole GS at $J=0.3$ (whose exact values are also shown on the plot).
}
\end{figure}

The 2D t--J Hamiltonian reads,
\begin{eqnarray}
H&=&  \sum_{i,j} J_{ij}({\bf
  S}_{i} \cdot {\bf S}_{j}-\frac{1}{4}n_{i}
n_{j}) \nonumber \\
&+& \sum_{i,j,\sigma} t_{ij}
({\tilde c}_{i,\sigma}^\dagger {\tilde c}_{j,\sigma} + h.c.) 
\label{Ham}
\end{eqnarray} 
where the exchange integrals $J_{ij}$ and hopping terms are (for simplicity)
restricted to nearest neighbor (NN) sites (called hereafter $J$ and $t$)
and ${\tilde c}_{i,\sigma}^\dagger$ are {\it projected} fermion
creation operators defined as $c_{i,\sigma}^\dagger (1-n_{i,-\sigma})$.
Hereafter, unless specified otherwise, $t$ sets the energy scale.
The conclusions drawn in this Letter are supported by Exact
Diagonalisation (ED) studies performed on the square cluster of N=32 sites   
depicted in Fig.~\ref{fig:cluster}(a) slightly doped with 
up to 2 holes~\cite{note1}.
This finite-size system is particularly appealing since it exhibits
the full local symmetries of the underlying square lattice
as well as all the most symmetric ${\bf k}$-points in reciprocal space
as seen in Fig.~\ref{fig:cluster}(b).
Note that, although the hole doping is quite small, the hole occupation
shown in Fig.~\ref{fig:cluster}(b) for a few $\bf k$-points 
is rather consistent with a ''large'' Fermi surface. 

A superconducting ground state (GS) is 
characterized by  Gorkov's
off-diagonal one-electron time-ordered Green function 
\begin{equation}
F_{\bf k}(t)=-\left\langle 
T {\tilde c}_{{\bf -k},-\sigma}(t/2)
{\tilde c}_{{\bf
k},\sigma}(-t/2)\right\rangle\, .
\label{eq:fk_of_tau}
\end{equation} 
Close to half-filling, in a finite system, 
$F_{\bf k}$ can be computed from,
\begin{equation}
{\tilde F}_{\bf k}(z)=\left\langle N-2 |{\tilde c}_{{\bf -k},-\sigma} 
\frac{1}{z-H+{\bar E}_{N-1}}   
{\tilde c}_{{\bf k},\sigma} | N
\right\rangle \, ,
\label{eq:fk_of_z}
\end{equation}
defined for all complex $z$ (with ${\rm Im}\,z\ne 0$).
Here the number of particles in the initial $| N \big>$ (half-filling)
and final $| N-2 \big>$ 
(two-hole doped) GS differ by two, reflecting charge fluctuations 
in a SC state.
For convenience, the energy reference ${\bar E}_{N-1}$ is defined as the
average between the GS energies of $| N \big>$ and $| N-2 \big>$.
Note that these states are both spin singlets.
In addition, they exhibit s-wave and d$_{x^2-y^2}$ orbital symmetries
(both of even parity) respectively. 
Owing to these special features, it is straightforward to show that
the analytical continuation of $F_{\bf k} (\omega)$ to the
real frequency axis is {\it even} in frequency 
and can be expressed as,
\begin{equation}
F({\bf k},\omega)={\tilde F}_{\bf k}(\omega+i\epsilon)
+{\tilde F}_{\bf k}(-\omega+i\epsilon)\, ,
\label{eq:fk_of_w}
\end{equation}
where $\epsilon$ is a small imaginary part. 
The superconducting frequency-dependent gap function 
$\Delta({\bf k},\omega)$ is directly proportional to $F({\bf k},\omega)$
and a simple analysis shows that both functions have 
d$_{x^2-y^2}$ orbital symmetry. In particular, they identically vanish 
on the diagonals $k_x=\pm k_y$.

\begin{figure}
\vspace{2mm} 
\includegraphics[width=0.45\textwidth]{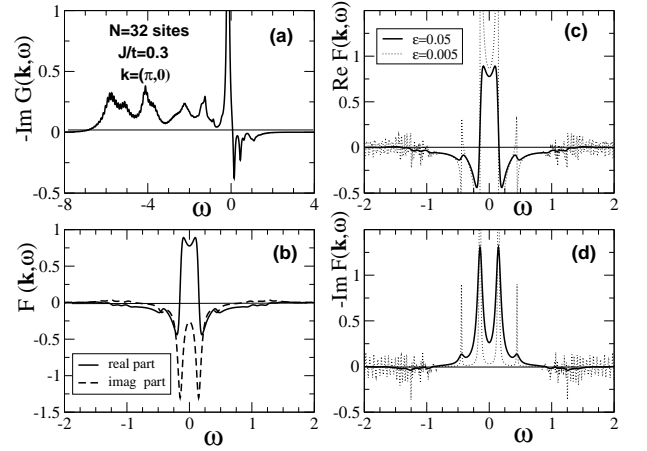}
\caption{\label{fig:dynamics}
Diagonal and off-diagonal one-electron (time-ordered) Green functions
versus frequency (in units of $t$) obtained on the cluster of 
Fig.~\protect\ref{fig:cluster} and an average hole doping of $3\%$;
(a) $-\rm{Im}\, G({\bf k},\omega)$; (b) real and imaginary parts of
$F({\bf k},\omega)$ for $\epsilon=0.05$; (c,d) comparison between 
the data of $\rm{Re}\, F$ and $-\rm{Im}\, F$ obtained with $\epsilon=0.05$ 
and with $\epsilon=0.005$.
}
\end{figure}

Pairing between holes should also be reflected in 
the structure of the (time-ordered)
diagonal Green function $G({\bf k},\omega)$. Here, it is convenient
to define the finite size $G({\bf k},\omega)$ as the sum of the
following electron- (i.e. occupied states for $\omega<0$) and hole-like 
(i.e. empty states for $\omega>0$) parts, 
\begin{eqnarray}
&&G({\bf k},\omega)=\left\langle N |{\tilde c}_{{\bf k},\sigma}^\dagger  
\frac{1}{\omega-i\epsilon+H-{\bar E}_{N-1}}
{\tilde c}_{{\bf k},\sigma} | N
\right\rangle \nonumber \\
&+&\big< N-2 |{\tilde c}_{{\bf k},\sigma} 
\frac{1}{\omega+i\epsilon-H+{\bar E}_{N-1}}   
{\tilde c}_{{\bf k},\sigma}^\dagger | N-2 \big> \, ,
\label{eq:gk_of_w}
\end{eqnarray}
so that both Green functions $F$ and $G$ 
have the {\it same} set of energy poles.
Here, the  well-known continued-fraction method used to 
compute {\it diagonal} correlation functions
such as $G({\bf k},\omega)$ is extended
to deal with {\it off-diagonal} ones 
such as 
$F({\bf k},\omega)$
\cite{OSEM94,continued_frac}.
Data are shown in Fig.~\ref{fig:dynamics}(a-d). The spectral density
$\rm{Im}\, G({\bf k},\omega)$ exhibits
sharp quasi-particle-like peaks
both above and below the Fermi level ($\omega=0$). 
Contrary to $G({\bf k},\omega)$, $F({\bf k},\omega)$ has a significant
amplitude only at low energy, typically for $|\omega|< 4J$ as seen in 
Fig.~\ref{fig:dynamics}(b), reflecting the energy scale of the pairing 
interaction. Note that, due to the discreteness of the low energy 
spectrum of the cluster, a finite value of $\epsilon$ is necessary to wipe out
the irrelevant fast oscillations of the $F$ (or $G$) Green functions
(see e.g. Fig.~\ref{fig:dynamics}(c-d)). However, the static limit
$\omega\rightarrow 0$ (together with $\epsilon \rightarrow 0$) is
perfectly controled and has a physical meaning.

\begin{figure}
\vspace{2mm} 
\includegraphics[width=0.43\textwidth]{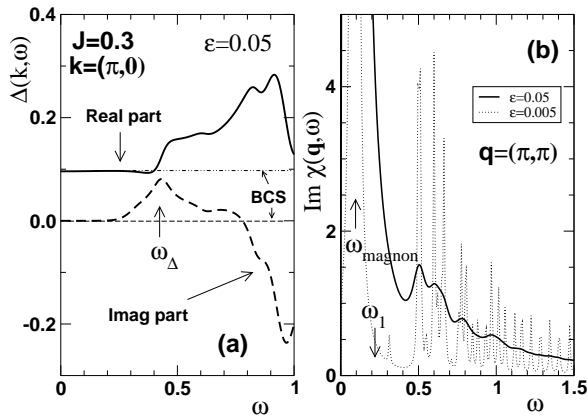}
\caption{(a) Real and imaginary components of the frequency-dependent gap 
function (in units of $t$) at low energies. 
(b) Dynamical spin structure factor in the 2 hole-doped 32-site cluster. 
The positions in energy of the first pole (magnon) and of the mean value
of the spectrum are indicated by arrows.
\label{fig:gap.modif} 
}
\end{figure}

The frequency-dependent gap function $\Delta({\bf k},\omega)$
can be extracted from the knowledge 
of the Green functions by assuming generic forms of a SC 
GS~\cite{schrieffer} {\it at low energies},
\begin{eqnarray}
G({\bf k},\omega)&=&z({\bf k},\omega)
\frac{\omega+\epsilon({\bf k},\omega)}{\omega^2-(\epsilon({\bf k},\omega)^2
+\Delta({\bf k},\omega)^2)+i\epsilon}, \nonumber \\
F({\bf k},\omega)&=&z({\bf k},\omega)
\frac{\Delta({\bf k},\omega)}{\omega^2-(\epsilon({\bf k},\omega)^2
+\Delta({\bf k},\omega)^2)+i\epsilon},
\label{eq:g&f}
\end{eqnarray}
where $z({\bf k},\omega)$ and $\epsilon({\bf k},\omega)$
are the inverse renormalization parameter and 
the renormalized energy 
(containing the energy shift) respectively.
In the BCS limit, the renormalized energy 
$\epsilon ({\bf k},\omega)\equiv\epsilon_{\bf k}$ and the gap function
$\Delta({\bf k},\omega)\equiv\Delta_{\bf k}$ do not depend on 
frequency (and $z({\bf k},\omega)=1$) so that 
$F({\bf k},\omega)$ exhibits only two poles at $\pm (\epsilon_{\bf k}^2
+\Delta_{\bf k}^2)^{1/2}$. 
However, it is necessary to assume an explicitely frequency dependence
of the gap function in order to reproduce the secondary peaks 
seen in Fig.~\ref{fig:dynamics}(d).
Using the parity of $\epsilon({\bf k},\omega)$ and $\Delta({\bf k},\omega)$
in the vicinity of the 
real frequency axis one gets,
\begin{equation}
\Delta({\bf k},\omega)=
\frac{2\omega F({\bf k},\omega)}{G({\bf k},\omega)-G({\bf k},-\omega)}
\label{eq:delta}
\end{equation}
Thus,
from a numerical calculation of $G({\bf k},
\omega)$ and $F({\bf k}, \omega)$, one can obtain
$\Delta({\bf k}, \omega)$.

Results for $\Delta ({\bf k}, \omega)$ with
${\bf k}=(\pi, 0)$ at the $M$ point, are shown in
Fig.~3(a).  Despite the limited resolution
in frequency of the ED data, $\Delta ({\bf k},
\omega)$ exhibits dynamic structure on a
scale of energies several times J.  We
believe that finite-size effects have
increased the onset frequency
$\omega_\Delta$ and that the corresponding
time $2\pi/\omega_\Delta$ should be viewed
as a lower bound of the characteristic time
scale of the pairing interaction.  For
comparison, the dynamic spin structure
factor $Im\chi ({\bf q}, \omega)$ for the 2-hole
doped 32-site cluster with ${\bf q}=(\pi, \pi)$ is
shown in Fig.~3(b). 

The quasi-particle gap at the gap edge,
where $\epsilon ({\bf k},0)=0$, is determined from the
self-consistent solution of
\begin{equation}
Re\Delta ({\bf k}, \omega = \Delta_{\bf k}) = \Delta_{\bf k}
\label{eight}
\end{equation}
As shown in Fig.~3(a), $\Delta({\bf k}, \omega)$
is real and essential constant over an
energy region larger than the gap.
Furthermore, as we will see, $\epsilon ({\bf k},0)$
for ${\bf k}=(\pi, 0)$ is small so that $\Delta_{\bf k}$
is to a good approximation
given by $\Delta({\bf k}, 0)$.  For ${\bf k}=(\pi,
0)$, the zero frequency limit of the gap
function $\Delta({\bf k}, \omega \to 0)$ (which is
purely real) computed by numerically taking
the $\omega$ and $\epsilon \to 0$ limits of
eq.~(\ref{eq:delta}) \cite {note2} is
plotted versus J in Fig.~4(a).  In
Fig.~4(b) the static $\omega=0$ value of the
renormalization energy  $\epsilon ({\bf k}, 0)$, the
renormalization factor $z({\bf k},\omega = 0)$, and the
equal-time pair amplitude $\langle
{\tilde c}^\dagger_{-{\bf k}, -\sigma} {\tilde c}^\dagger_{{\bf k}\sigma}
\rangle$ for ${\bf k}=(\pi, 0)$ versus J are also
shown \cite{note2_bis}. The values of
$\Delta (k,0)$ and $z(k,0)$ are in good
agreement with the results Ohta, et.~al
\cite{OSEM94} 
found by fitting their numerical t-J
$4\times 4$ and $\sqrt{18}\times
\sqrt{18}$ cluster calculations to a BCS
quasi-particle form in which the
frequency dependence of $\Delta$ and $z$
were neglected.  They found that the
magnitude of the gap at the antimode
varies as 0.25J to 0.3J at low doping
and as seen in Fig.~4(a) we find $\Delta
(k,0) \sim 0.3J$.  It is interesting to
note that the data for $\Delta (k,0)$ in
Fig.~4(a) do not show any
lower bound of J in contrast to the pair binding 
energy \cite{Poi93,Leu00}. 
The spin structure factor is plotted in 
Fig.~3(b)
showing a low energy peak (magnon) and a higher energy background 
that could be connected to
the structures in the gap function at an energy $\sim J$. 
For comparison, the magnon energy and 
the mean-value~\cite{note3} of the spin spectral weight 
are plotted in Fig.~4(a) together with the 
zero-frequency gap
and the onset frequency $\omega_\Delta$ in the gap function.
While the
similarities in the J-dependence
of these quantities suggest that the
dynamics underlying the pairing mechanism is
related to the spin fluctuations, 
further work on other clusters, such as
2-leg ladders, are needed to sort out the
influence of finite-size effects.
\begin{figure}
\vspace{2mm}
\includegraphics[width=0.45\textwidth]{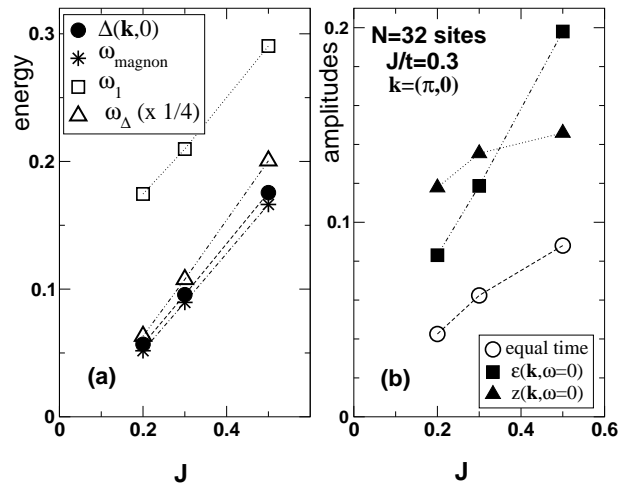}
\caption{(a) Zero-frequency gap and typical
frequency $\omega_\Delta$ (divided by 4) of
the gap function (in units of $t$) plotted
vs J.  For comparison, the magnon energy and
the mean-value $\omega_1$ of the spin
structure factor (see Fig.~3(b) is also
shown.
(b) Equal-time pair amplitude $\langle
{\tilde c}^\dagger_{-{\bf k}, -\sigma} 
{\tilde c}^\dagger_{{\bf k},\sigma}\rangle$ calculated at momentum
${\bf k}=(\pi, 0)$ vs $J/t$.  The renormalized energy
$\epsilon({\bf k}, 0)$ (in units of $t$) 
and the renormalization
parameter $z({\bf k},\omega=0)$ are also shown as
filled squares and triangles, respectively. 
\label{fig:gap.modif}
}
\end{figure}

\begin{table}
\caption{Static limit ($\omega=0$) of the gap function, the renormalized energy
and the renormalization parameter at a few ${\bf k}$-points away from the nodes
and for $J/t=0.3$.}
\begin{center} 
\begin{tabular}{|c|c|c|c|}
\hline
	&	&	& \\
 {\hskip 0.4cm $ {\bf k}$-points \hglue 0.4cm} 
&  {\hskip 0.4cm $\Delta({\bf k},0)$ \hskip 0.4cm }
&  {\hskip 0.4cm $\epsilon ({\bf k},0)$ \hskip 0.4cm }
&  {\hskip 0.4cm $z ({\bf k},0)$ \hskip 0.4cm} \\
	&	&	& \\ \hline
	&	&	& \\ 
 $(\pi/2,0)$   & 0.064958 \hfill& 0.20260  & 0.21221 \\
 $(\pi,0)$     & 0.095641 & 0.11867  & 0.13542  \\
 $(\pi,\pi/2)$ & 0.186100 & 0.08895  & 0.12855  \\
	&	&	& \\ \hline
\end{tabular}
\end{center}
\label{table_phd}
\end{table}

We conclude this investigation with a
further discussion of the parameters which
enter eq.~(\ref{eq:g&f}). Note that since
these parameters depend upon both ${\bf k}$ and
$\omega$, the static $\omega=0$ results
which we will discuss are near the energy
shell only for ${\bf k}$ values near the fermi
surface and for $\Delta_{\bf k}$ small compared to
the characteristic frequency variations set
by J.  In fact, even trying to select
${\bf k}$-points near the ``Fermi surface''
defined by $\epsilon ({\bf k}, 0)=0$ requires some
care since the Fermi surface (FS) cannot be
exactly defined on a finite cluster,
especially for a small number of holes.
However, the smooth variation of the hole
occupancy in the BZ [see Fig.~1(b)] suggests
that this system should still pick up
features of a slightly doped AF with a large
FS.  With this in mind, the static limit of
the renormalized energy $\epsilon ({\bf k}, 0)$
and the renormalization parameter $z({\bf k},0)$
were obtained from the calculated $\epsilon,
\omega \to 0$ limits of $G({\bf k}, 0)$, $F({\bf k},
0)$, and $\Delta ({\bf k}, 0)$ by using
eq.~(\ref{eq:g&f}) for $\omega=0$.  These
are tabulated for values of ${\bf k}$ both
near the nominal FS and away from it in
Table 1.  The renormalized energy
$\epsilon({\bf k}, 0)$ has roughly the energy
scale of the quasi-particle-like energy
poles at the available momenta of the BZ.
The small value of $\epsilon({\bf k}, 0)$ given by
our analysis for ${\bf k}=(\pi, 0)$ show that the
excitation that we are probing is relatively
close to the FS (defined here as having
$\epsilon({\bf k}, 0)=0$.  Note also that the
weight $z({\bf k},0)$ of these excitations is
small $(\approx J/2)$ compared to the large
incoherent background [see Fig.~2(a)]. Its
J-dependence plotted in Fig.~4(b) is
consistent with the $J^\alpha$ power-law
behavior found previously in earlier
calculations of a single hole propagating in
an AF background \cite{PSZ93}.

\acknowledgments

D.J.~Scalapino would like to acknowledge support from the 
US Department of Energy under Grant No DE-FG03-85ER45197.
D.~Poilblanc thanks M.~Sigrist (ETH-Z\"urich) for 
discussions and aknowledges hospitality of the Physics Department 
(UC Santa Barbara) where part of this work was carried out. 
Numerical computations were done on the vector NEC-SX5 supercomputer
at IDRIS (Paris, France).


\begin{references}
\bibitem{continued_frac}  Further
details of our calculation will be published 
elsewhere.
\end{references}
\end{document}